\documentstyle[prb,aps,psfig,epsf,floats]{revtex}
\newcommand{\beq}{\begin{equation}}
\newcommand{\eeq}{\end{equation}}
\newcommand{\beqn}{\begin{eqnarray}}
\newcommand{\eeqn}{\end{eqnarray}}

\newcommand{\bfi}{\begin{figure}}
\newcommand{\efi}{\end{figure}}
\newcommand{\var}[2]{{\mbox{}_{#2} \atop {\bar #1}}}

\begin{document}
\title{\bf Violation of Luttinger's theorem in strongly correlated 
electronic systems within a 1/N expansion }

\vspace{3cm}
\author{{Emmanuele Cappelluti} and {Roland Zeyher}}

\address{Max-Planck-Institut\  f\"ur\
Festk\"orperforschung,\\ Heisenbergstr.1, 70569 Stuttgart, Germany \\}

\date{\today }

\vspace{3cm}

\maketitle

\begin{abstract}

We study the $1/N$ expansion of a generic, strongly correlated electron
model ($SU(N)$ symmetric Hubbard model with $U=\infty$ and $N$ degrees of
freedom per lattice site) in terms of $X$ operators. The
leading order of the expansion describes a usual Fermi liquid with 
renormalized, stable particles. The next-to-leading
order violates 
Luttinger's theorem if a finite convergence radius for the
$1/N$ expansion for a fixed and non-vanishing doping away from 
half-filling is assumed. We find that the  
volume enclosed by the Fermi surface, is at large, but finite $N$'s
and small dopings larger than at $N=\infty$. 
As a by-product an explicit expression for the electronic
self-energy in $O(1/N)$ is given which cannot be obtained by factorization
or mode-coupling assumptions but contains rather sophisticated vertex
corrections.
\end{abstract}

\section{Introduction}

Luttinger's theorem \cite{Lutt1,Lutt2} is one of the most fundamental 
theorems in solid
state physics. It states that the volume enclosed by the Fermi
surface is independent of the strength of the interaction between
electrons. One definition of the Fermi surface is based on the
momentum distribution function $n({\bf k})$ of electrons. The
Fermi surface is the set of all ${\bf k}$-points where $n({\bf k})$
or derivatives of it are singular. If the imaginary part of the 
self-energy $\Sigma({\bf k},\omega)$ of the one-particle Green's function 
$G({\bf k},\omega)$ vanishes at $\omega = 0$ 
the Fermi surface can also be defined as the boundary of the 
set of $\bf k$ points satisfying $Re G({\bf k},\omega = 0) > 0$.
If the electrons form a Fermi liquid $n({\bf k})$ possesses a jump
and $Im \Sigma({\bf k},\omega) \sim \omega^2$ near the Fermi surface.
Both definitions for the Fermi surface are then equivalent and Luttinger's
theorem also holds. Fermi liquid behavior of the electrons, however, is not
a necessary prerequisite for the validity of Luttinger's theorem:
One-dimensional metals are Luttinger and not Fermi liquids, yet
Luttinger's theorem also holds in this case \cite{Bedell,Affleck}. However, 
there is 
evidence that Luttinger's theorem may be violated in strongly correlated
electronic systems \cite{Putikka,Eder}. 
Based on high temperature expansions for the
two-dimensional $t-J$ model it was concluded that Luttinger's theorem
is violated in this case even at a doping as large as 20 per 
cent\cite{Putikka}. In Ref.\cite{Eder} finite temperature Quantum Monte-Carlo
simulations for the Hubbard model with $U/t=8$ and $T/t=0.33$ ($U$
and $T$ are the Hubbard repulsion and the temperature, respectively)
yielded a doping dependence for the volume enclosed by the Fermi surface
which was similar to that of the Hubbard I approximation and thus much
larger near half-filling than that predicted by the Luttinger theorem.

Theories based on perturbation expansions in powers of the interaction
between electrons admit non-perturbative approximations which fulfill
Luttinger's theorem \cite{Baym}. Examples of this kind are all conserving 
approximations. In strongly correlated electronic systems the interaction 
energy
dominates over the kinetic energy and expansions around the atomic limit
in terms of the kinetic energy seem appropriate. Various general formulations
of such expansions have been given \cite{Kotliar,Metzner,Ruckenstein,Zeyher1}.
It seems, however, that so far 
no well-defined approximation within such an expansion has been found 
for a finite dimension which explicitly obeys Luttinger's theorem.
Approximate treatments based on mode-coupling \cite{Prelovsek} or 
factorization assumptions \cite{Plakida}
seriously violate Luttinger's theorem. Since the employed approximations
are uncontrolled it is not possible to judge whether these violations
describe real effects or represent just artifacts.

The use of $1/N$ expansions is one way to create controlled 
approximations for strongly correlated systems. Formally, one extends
the $SU(2)$ symmetry group for the spins in the Hamiltonian to a
larger symmetry group $SU(N)$ or $Sp(N/2)$ and considers $1/N$ as a
small expansion parameter. The extension to $N$ degrees of freedom 
is not unique, the two most popular alternatives are the slave boson or
the slave fermion theories. The $1/N$ expansion can also be
carried out directly in terms of $X$ operators, the analogue of the
slave boson extension has been worked out in detail in Refs.
\cite{Zeyher1,Kulic1,Zeyher2,Kulic2}. The 
$1/N$ expansion allows to study the validity of Luttinger's 
theorem as a power expansion in $1/N$. If this theorem holds for $N=2,3...$
and if the convergence radius of the $1/N$ expansion is non-zero the
theorem must hold in each order in $1/N$. It is easy to show that
the theorem holds for $N = \infty$. In the following we will calculate
the change in the Fermi surface $\sim 1/N$ and from this the corresponding
change $\sim 1/N$ of the volume enclosed by the Fermi surface. In this
way the validity of Luttinger's theorem for strongly correlated 
systems can be assessed under the only assumption that the $1/N$ expansion
has a finite convergence radius. Explicit results are presented for the $t-J$
model with $J=0$, the so-called $t$ model, given, for instance, by 
the first term on the right-hand side of Eq.(1) of Ref.\cite{Zeyher2}.
One main ingredient of the calculation is the 
electron self-energy calculated in order $1/N$. Below we describe
a more symmetric and transparent derivation of this quantity compared
to that in Ref.\cite{Kulic2} which also allows to determine the 
$N$-dependence of
the various contributions in a straightforward way. We also will 
discuss the change in the $n(\mu)$ relation ($n$ is the number of electrons
per site, $\mu$ the chemical potential) if the self-energy contribution
of $O(1/N)$ with $N=2$ is used.

\section{ Electronic self-energy in $O(1/N)$}

In this section we calculate the contribution of $O(1/N)$ to the electronic 
self-energy $\Sigma$. It represents the first correction beyond the mean-field
approximation for $\Sigma$. We consider the $t-J$ model and use the
general framework outlined in Ref.\cite{Zeyher1}. Fermionic Hubbard 
operators, which
annihilate or create an electron, will be denoted by $X_e(1)$ and $X_h(1)$,
respectively. Writing out the arguments explicitly we would have 
$X^{0p}(i_1\tau_1)$
and $X^{p0}(i_1\tau_1)$, respectively, where $p=1,...,N$ denotes an
internal index and $i_1,\tau_1$ a lattice site and an imaginary time,
respectively. The remaining bosonlike Hubbard operators will often be
denoted by the letter $Y$. $Y(1)$ stands for $Y^{p_1q_1}(i_1\tau_1)$
with $p_1,q_1=0,...,N$ with the cases $p_1=0,q_1>0$ and $p_1>0,q_1=0$
excluded. It will also be often convenient to abbreviate the
index pair $i_1\tau_1$ by ${\bar 1}$. All explicitly given formulas
in the following refer to the $t-J$ model with $J=0$. Using the rules
discussed in Ref.\cite{Zeyher2} it is straightforward to generate from them
the corresponding contributions due to a spin-spin or a 
charge-charge interaction term. 

We start from the equations of motion for X operators,
\beq
\frac{\partial X_e(1)}{\partial \tau_1} =
\int d3 \mu_e(13)X_e(3)+
\int d2 d3 t_e(123)Y(2)X_e(3),
\label{emot}
\eeq
\beq
\frac{\partial X_h(1)}{\partial \tau_1} =
\int d3 \mu_h(13)X_h(3)+
\int d2 d3 t_h(123)Y(2)X_h(3).
\label{hmot}
\eeq
The functions $\mu$ and $t$ in Eqs. (\ref{emot}) and (\ref{hmot}) are 
defined by
\beq
\mu_e(12)=\mu \delta(\bar{1}-\bar{2}) \delta_{q_1 q_2},
\label{mue}
\eeq
\beq
\mu_h(12)=-\mu \delta(\bar{1}-\bar{2}) \delta_{p_1 p_2},
\label{muh}
\eeq
\beq
t_e(123)=\frac{t(\bar{1}-\bar{3}^-)}{N} 
\delta(\bar{1}-\bar{2})
[\delta_{p_2 0} \delta_{q_2 0} \delta_{q_1 q_3} +
\delta_{q_1 q_2} \delta_{p_2 q_3}],
\label{te}
\eeq
\beq
t_h(123)=-\frac{t(\bar{3}-\bar{1}^-)}{N} 
\delta(\bar{1}-\bar{2})
[\delta_{p_2 0} \delta_{q_2 0} \delta_{p_1 p_3} +
\delta_{p_1 p_2} \delta_{q_2 p_3}].
\label{th}
\eeq
The electronic Green's function $G$ is defined by
\beq
G(12) = - \langle T X_e(1) X_h(2) \rangle,
\eeq
where $T$ is the time ordering operator. 
Dyson's equation for $G$ reads:
\beq
G^{-1}(12) = \int d3 Q^{-1}(13)\left[
G_0^{-1}(32)+\mu_e(32)-\Sigma_e(32)
\right].
\label{dysinvert}
\eeq
The free Green's function $G_0^{-1}$ is given by
\beq
G_0^{-1}(12) = - \delta(\bar{1}-\bar{2})
[\delta_{p_1 0} \delta_{p_2 0} \delta_{q_1 q_2}+
\delta_{q_1 0} \delta_{q_2 0} \delta_{p_1 p_2}]
\frac{\partial}{\partial \tau_2}.
\eeq
There is no need to specify any electron or hole index in $G_0^{-1}$
since $G_0^{-1}$ is identical in both cases. $Q$ is the spectral
weight given by the expectation value of the equal-time anticommutator
of the two fermionic operators of $G$, i.e., writing out explicitly the
internal indices, by
\beq
Q({{0p}\atop{\bar 1}} {{q0}\atop{\bar 2}}) = \delta ({\bar 1}- {\bar 2})
(\langle Y^{00}({\bar 1}) \rangle \delta_{pq} + \langle Y^{qp}({\bar 1})
\rangle ).
\eeq 
The self-energy $\Sigma_e(12)$ is given by 
\beq
\Sigma_e(12) = \int d3d4d5 t_e(134) \langle T Y(3) X(4) X(5)\rangle
G^{-1}(52).
\label{seinvert}
\eeq
Using Eq.(\ref{dysinvert})
we apply the operator $G^{-1}$ in Eq.(\ref{seinvert}) to the 
left and act with it on the time-ordered product of Hubbard operators.
Using Eqs.(\ref{emot})-(\ref{muh})
and evaluating equal-time anticommutators we obtain
\beqn
\Sigma_e(12)&=&-\int d3d4d5 t(1345) \langle Y(3) Y(4)\rangle Q^{-1}(52)
+\int d3d4d5 \lambda(1345) \langle X_h(3) X_e(4)\rangle Q^{-1}(52)\nonumber\\
&&+ \int d3d4d5d6d7 t_e(134) t_h(756) \langle T Y(3) Y(5) X_e(4) X_h(6)\rangle
Q^{-1}(72),
\label{selung2}
\eeqn
with
\beqn
t(1342)&=&\frac{t(\bar{1}-\bar{2})}{N} 
\delta(\bar{1}-\bar{3})\delta(\bar{2}-\bar{4})
[\delta_{q_1 p_2} \delta_{p_3 0} \delta_{q_3 0} \delta_{p_4 0} \delta_{q_4 0} 
+\delta_{q_1 q_4} \delta_{p_4 p_2} \delta_{p_3 0} \delta_{q_3 0}
\nonumber\\
&&
+\delta_{q_1 q_3} \delta_{p_3 p_2} \delta_{p_4 0} \delta_{q_4 0}
+\delta_{q_1 q_3} \delta_{p_3 q_4} \delta_{p_4 p_2}],
\eeqn
and
\beq
\lambda(1342)=\frac{t(\bar{1}-\bar{4})}{N} 
\delta(\bar{1}-\bar{2})\delta(\bar{1}-\bar{3})
[\delta_{q_1 p_2} \delta_{p_3 q_4} - \delta_{q_1 q_4} \delta_{p_3 p_2}].
\eeq 
In Eq.(\ref{selung2}) we used the fact that reducible self-energy
contributions cancel each other and thus can be dropped. Since this
property of self-energy diagrams is not completely obvious in our
case we give a proof of it in appendix A. The first two
terms in Eq.(\ref{selung2}) are frequency-independent and thus
contribute only to the dispersion of quasi-particles. The third term
in Eq.(\ref{selung2}) depends on the frequency and determines the
damping of the quasi-particles.

It is convenient to introduce an external source field $K$ coupled to bosonic
Hubbard operators for the evaluation of the expectation values in 
Eq.(\ref{selung2}) similar as in Ref.\cite{Zeyher1}. 
The self-energy assumes then the form
\beqn
\Sigma_e(12) &=&
- \int d3 t_e(132) \langle Y(3)\rangle - 
\int d3d4d5 t(1345) 
\frac{\delta \langle Y(4) \rangle}{\delta K(3)} Q^{-1}(52)\nonumber\\
&&+\int d3d4d5 \lambda(1345) \langle X(3) X(4) \rangle Q^{-1}(52)\nonumber\\
&& + \int d3d4d5d6d7d8d9 t_e(134) t_h(956)
G(47) \frac{\delta \Gamma(78;5)}{\delta K(3)} G(86)Q^{-1}(92)\nonumber\\
&& - \int d3d4d5d6d7d8d9d10d11 t_e(134) t_h(1156)
G(47)\Gamma(78;5)G(89)\Gamma(910;3)G(106)Q^{-1}(112)\nonumber\\
&& - \int d3d4d5d6d7 t_e(134) t_h(756)
\frac{\delta \langle Y(5)\rangle}{\delta K(3)} G(46)Q^{-1}(72).
\label{dersp}
\eeqn
$\Gamma$ is a vertex function defined by
\beq
\Gamma(12;3) = \frac{\delta G^{-1}(12)}{\delta K(3)}.
\label{Gamma}
\eeq
We tacitly assumed in Eq.(\ref{dersp}), and will do so also in the following,
that the external source $K$ is put to zero once the functional derivatives
have been carried out. The evaluation of the fourth term in 
Eq.(\ref{dersp}) is somewhat more involved than that of the other
terms and is therefore deferred to appendix B. In the following we often will
abbreviate this term by $\delta \Sigma$.
Summing over the internal indices or exhibiting them
explicitly Eq.(\ref{dersp}) becomes
\beqn
&&  \Sigma^{\sigma \sigma}(\bar{1}\bar{2}) =
- \frac{t(\bar{1}-\bar{2})}{N} \left[ \langle Y^{00}\rangle
+ \langle Y^{\sigma \sigma}\rangle \right]
+(N-1) \frac{t(\bar{1}-\bar{3})}{N} g(\bar{3}\bar{1}^+)
\delta(\bar{1}-\bar{2})  \nonumber\\
&& +(\frac{t(\bar{1}-\bar{3})}{N}g(\bar{3}\bar{4})\frac{t(\bar{4}
-\bar{2})}{N}
- \frac{t(\bar{1}-\bar{2})}{N [\langle Y^{00}\rangle
+ \langle Y^{\sigma \sigma}\rangle ]}) \times \nonumber\\
&& \left[
- \frac{\delta \langle Y^{\sigma_1 \sigma_1}(\bar{2}) \rangle}
{\delta K^{00}(\bar{1})}
+ \frac{\delta \langle Y^{\sigma \sigma}(\bar{2}) \rangle}
{\delta K^{00}(\bar{1})}
- \frac{\delta \langle Y^{\sigma_1 \sigma_1}(\bar{2}) \rangle}
{\delta K^{\sigma \sigma}(\bar{1})}
+ \frac{\delta \langle Y^{\sigma \sigma_1}(\bar{2}) \rangle}
{\delta K^{\sigma_1 \sigma}(\bar{1})}\right] \nonumber\\
&&+ \frac{t(\bar{1}-\bar{3})}{N}g(\bar{3}\bar{4})g(\bar{5}\bar{6})
g(\bar{7}\bar{8})\frac{t(\bar{8}-\bar{2})}{N}
[\langle Y^{00}\rangle + \langle Y^{\sigma \sigma}\rangle]^2
[
\Gamma^{\sigma\sigma}_{00}(\bar{4}\bar{5};\bar{2})
\Gamma^{\sigma\sigma}_{00}(\bar{6}\bar{7};\bar{1})
\nonumber\\
&&
+\Gamma^{\sigma_1\sigma_1}_{00}(\bar{4}\bar{5};\bar{2})
\Gamma^{\sigma_1\sigma}_{\sigma_1\sigma}(\bar{6}\bar{7};\bar{1})
+\Gamma^{\sigma\sigma_1}_{\sigma\sigma_1}(\bar{4}\bar{5};\bar{2})
\Gamma^{\sigma_1\sigma_1}_{00}(\bar{6}\bar{7};\bar{1})
-\Gamma^{\sigma\sigma}_{\sigma\sigma}(\bar{4}\bar{5};\bar{2})
\Gamma^{\sigma\sigma}_{\sigma\sigma}(\bar{6}\bar{7};\bar{1})
\nonumber\\
&&
+\Gamma^{\sigma_1\sigma_1}_{\sigma\sigma}(\bar{4}\bar{5};\bar{2})
\Gamma^{\sigma_1\sigma}_{\sigma_1\sigma}(\bar{6}\bar{7};\bar{1})
+\Gamma^{\sigma\sigma_1}_{\sigma\sigma_1}(\bar{4}\bar{5};\bar{2})
\Gamma^{\sigma_1\sigma_1}_{\sigma\sigma}(\bar{6}\bar{7};\bar{1})
] +\delta \Sigma^{\sigma\sigma}({\bar 1}{\bar 2}).
\label{selung4}
\eeqn
In Eq.(\ref{selung4}) we have renamed the internal indices $p,p_1...$ by
$\sigma,\sigma_1...$ where $\sigma,\sigma_1...$ assume only values between 
1 and N.
We also used the fact that in equilibirum $\Sigma$ and $g$ are
diagonal and independent of spin indices, $Y$ is diagonal
and in $\Gamma$ either the rows or the columns have the same spin indices
(the upper two indices in $\Gamma$ result from the two fermionic, the
lower two indices from the bosonic Hubbard operators). Spin indices on the 
right-hand side of Eq.(\ref{selung4}) which do not appear on the left-hand
side of this equation are summed over. In accordance with that 
we explicitly show the index $\sigma$, for instance,
at the self-energy, in spite of the fact that $\Sigma$ is independent of 
spin labels in equilibrium. In Eq.(\ref{selung4}) we also have introduced
a normalized Green's function by means of
\beq
G(12) = \int d3 g_e(13)Q(32) = \int d3 Q(13)g_h(32).
\label{gedef}
\eeq
In equilibirum we have $g_e=g_h$ so we simply can use the letter
$g$ in Eq.(\ref{selung4}) for convenience. Let us introduce normalized vertex
functions by
\beq
\gamma_e(12;3)=\frac{\delta 
g_e^{-1}(12)}{\delta K(3)},
\eeq
\beq
\gamma_h(12;3)=\frac{\delta 
g_h^{-1}(12)}{\delta K(3)}.
\eeq
Unlike $g_e$ and $g_h$, $\gamma_e$ is not equal to $\gamma_h$
in equilibrium. The previously defined vertex function $\Gamma$
can now be expressed as
\beqn
\Gamma(12;3) &=&
\int d4 Q^{-1}(14)\gamma_e(42;3)+
\int d4 \frac{\delta Q^{-1}(14)}{\delta K(3)} g_e^{-1}(42)
\nonumber\\
&=&
\int d4 \gamma_h(14;3) Q^{-1}(42)+
\int d4 g_h^{-1}(14) \frac{\delta Q^{-1}(42)}{\delta K(3)}.
\label{vertex}
\eeqn
We stress that the two expressions for $\Gamma$ in terms of $e$- or 
$h$- quantities
are identical, so that we may take the more convenient one of the two 
whenever we use Eq.(\ref{vertex}).

Let us introduce some short notations that
will be useful later:
\beq
\Pi\left(\var{2}{\sigma_1\sigma_2};\var{1}{\sigma_3\sigma_4}\right)
=\frac{\delta \langle Y^{\sigma_1 \sigma_2}(\bar{2}) \rangle}
{\delta K^{\sigma_3 \sigma_4}(\bar{1})},
\eeq
\beq
\Gamma\left(\var{1}{\sigma_1}\var{2}{\sigma_2};
\var{3}{\sigma_3\sigma_4}\right)=
\Gamma^{\sigma_1\sigma_2}_{\sigma_3\sigma_4}(\bar{1}\bar{2};\bar{3}),
\eeq
\beq
\gamma\left(\var{1}{\sigma_1}\var{2}{\sigma_2};
\var{3}{\sigma_3\sigma_4}\right)=
\gamma^{\sigma_1\sigma_2}_{\sigma_3\sigma_4}(\bar{1}\bar{2};\bar{3}).
\eeq

The $1/N$ expansion can be simplified by noting that
\beq
O\left(
\left.\frac{{\delta A}^{\alpha\beta}}{{\delta K}^{\gamma\delta}}
\right|_{K=0}\right) \leq
O\left(\left.A^{\alpha\beta}\right|_{K=0}\right),
\label{order}
\eeq
where $A^{\alpha\beta}$ is a functional of $g$. $O(...)$ on the left- and
right-hand sides of Eq.(\ref{order}) denotes always the largest order
occurring in one of the tensor elements.
The functional derivative on the left-hand side of Eq.(\ref{order})
acts on a Green's function $g^{\sigma_1\sigma_2}$ and can be written as
\beq
{{\delta g^{\sigma_1\sigma_2}}\over {\delta K^{\sigma_3\sigma_4}}}
= -g^{\sigma_1\sigma_1}\gamma^{\sigma_1\sigma_2}_{\sigma_3\sigma_4}
g^{\sigma_2\sigma_2},
\label{abl}
\eeq
omitting space and time labels, zeros in fermionic 
index pairs and taking finally all quantities in equilibrium.
If $\sigma_3=\sigma_4$ in $K$ the number of internal sums on the two sides
of Eq.(\ref{order}) is equal. Since the orders $O(g)$ and $O(\gamma)$
are at most 1 Eq.(\ref{order}) holds. If $\sigma_3 \neq \sigma_4$ in $K$
the left-hand side of Eq.(\ref{order}) may contain first one sum more than
the right-hand side of this equation. However, because of Eq.(\ref{abl})
and because of the selection rules for $\gamma$, this additional sum does
not survive when finally going to equilibrium so that Eq.(\ref{order})
again holds. Using Eq.(\ref{order}) the expansion of various quantities 
in powers of $1/N$ read
\beq
\Sigma(\bar{1}\bar{2}) = \sum_{i=0}^{\infty}
\Sigma_i(\bar{1}\bar{2}),
\label{expansion1}
\eeq
\beq
g(\bar{1}\bar{2}) = \sum_{i=0}^{\infty}
g_i(\bar{1}\bar{2}),
\eeq
\beq
\langle Y^{00} \rangle =
\sum_{i=-1}^{\infty} \langle Y^{00} \rangle_i,
\eeq
\beq
\langle Y^{\sigma \sigma} \rangle =
\sum_{i=0}^{\infty} \langle Y^{\sigma \sigma} \rangle_i,
\eeq
\beq
\Pi(\bar{2};\bar{1}) =
\sum_{i=0}^{\infty}\Pi_i(\bar{2};\bar{1}),
\eeq
\beq
\Gamma(\bar{1}\bar{2};\bar{3}) =
\sum_{i=1}^{\infty}\Gamma_i(\bar{1}\bar{2};\bar{3}),
\eeq
\beq
\gamma(\bar{1}\bar{2};\bar{3}) =
\sum_{i=0}^{\infty}\gamma_i(\bar{1}\bar{2};\bar{3}).
\label{expansion2}
\eeq
Carrying out the $1/N$ expansion, we find for the leading order of 
the self-energy:
\beq
\Sigma_0^{\sigma\sigma}(\bar{1}\bar{2}) =
-\frac{\langle Y^{00} \rangle_{-1}}{N}t(\bar{1}-\bar{2})
+t(\bar{1}-\bar{3})g_0(\bar{3}\bar{1}^+)\delta(\bar{1}-\bar{2}).
\label{s0k}
\eeq
Eq.(\ref{s0k}) reads in Fourier space
\beq
\Sigma_0({\bf k}) =
-\frac{\langle Y^{00} \rangle_{-1}}{N}t({\bf k})
+\lambda_0,
\eeq
where
\beq
\lambda_0 = \sum_k t({\bf k}) g_0(k)
e^{i\omega_n 0^+},
\label{eqlambda}
\eeq
and
\beq
g_0({\bf k},i\omega_n) = \frac{1}{i\omega_n+ (\langle Y^{00} \rangle_{-1} / N)
t({\bf k})+\mu-\lambda_0}.
\label{g0}
\eeq

In Eq.(\ref{eqlambda}) and in the following equations 
the short notation $\sum_k$ stands for $(T/V) \sum_{{\bf k},\omega_n}$
where $T$ is the temperature and $V$ the volume
of the crystal. 

For the order $O(1/N)$ of $\Sigma$ we obtain:
\beqn
\Sigma_1^{\sigma\sigma}(\bar{1}\bar{2}) &=&
-\frac{t(\bar{1}-\bar{2})}{N}[\langle Y^{00} \rangle_{0}
+\langle Y^{\sigma\sigma} \rangle_{0}]
+t(\bar{1}-\bar{3})\left[
g_1(\bar{3}\bar{1}^+) - \frac{g_0(\bar{3}\bar{1}^+)}{N}\right]
\delta(\bar{1}-\bar{2})
\nonumber\\
&&+(
\frac{t(\bar{1}-\bar{2})}{\langle Y^{00} \rangle_{-1}}
-\frac{t(\bar{1}-\bar{3})}{N}g_0(\bar{3}\bar{4})t(\bar{4}-\bar{2}))
\left[
\Pi_0\left(\var{2}{\alpha\alpha};\var{1}{00}\right)+
\Pi_0\left(\var{2}{\alpha\alpha};\var{1}{\sigma\sigma}\right)-
\Pi_0\left(\var{2}{\sigma\alpha};\var{1}{\alpha\sigma}\right)
\right]
\nonumber\\
&&+
\frac{t(\bar{1}-\bar{3})}{N}
g_0(\bar{3}\bar{4})g_0(\bar{5}\bar{6})g_0(\bar{7}\bar{8})
t(\bar{8}-\bar{2})\langle Y^{00} \rangle_{-1}^2
\times
\nonumber\\
&&
\left[
\Gamma_1\left(\var{4}{\alpha}\var{5}{\alpha};\var{2}{00}\right)
\Gamma_1\left(\var{6}{\alpha}\var{7}{\sigma};\var{1}{\alpha\sigma}\right)
+\Gamma_1\left(\var{4}{\sigma}\var{5}{\alpha};\var{2}{\sigma\alpha}\right)
\Gamma_1\left(\var{6}{\alpha}\var{7}{\alpha};\var{1}{00}\right)
\right.
\nonumber\\
&&\left.
+\Gamma_1\left(\var{4}{\alpha}\var{5}{\alpha};\var{2}{\sigma\sigma}\right)
\Gamma_1\left(\var{6}{\alpha}\var{7}{\sigma};\var{1}{\alpha\sigma}\right)
+\Gamma_1\left(\var{4}{\sigma}\var{5}{\alpha};\var{2}{\sigma\alpha}\right)
\Gamma_1\left(\var{6}{\alpha}\var{7}{\alpha};\var{1}{\sigma\sigma}\right)
\right] +\delta \Sigma_1^{\sigma\sigma}({\bar 1}{\bar 2}).
\label{selung5}
\eeqn

We are now evaluating analytically all the unknown quantities in 
Eq.(\ref{selung5}) using the relation\cite{Zeyher1}
\beq
\left.\langle Y^{\sigma\sigma}(\bar{1}) \rangle\right|_{K=0} = 
g\left(\var{1}{\sigma}\var{1^+}{\sigma}\right)
-\Pi\left(\var{1}{\sigma\sigma};\var{1}{00}\right)
\frac{1}{\langle Y^{00} \rangle}+
\left[g\left(\var{1}{\sigma}\var{1^+}{\sigma}\right)-1\right]
\frac{\langle Y^{\sigma\sigma} \rangle}{\langle Y^{00} \rangle},
\label{in1}
\eeq
and the constraint
\beq
Y^{00}+\sum_{\sigma=1}^N Y^{\sigma\sigma} = \frac{N}{2}.
\label{constr}
\eeq
Performing explicitely the $1/N$ expansion, we obtain
\beq
\langle Y^{00} \rangle_{-1} = \frac{N}{2} - 
N \langle Y^{\sigma\sigma} \rangle_0,
\label{y00}
\eeq
\beq
\langle Y^{00} \rangle_{0} = - 
N \langle Y^{\sigma\sigma} \rangle_1,
\eeq
\beq
\langle Y^{\sigma\sigma} \rangle_0 = 
g_0\left(\bar{1}\bar{1}^+\right),
\label{y0}
\eeq
\beq
\langle Y^{\sigma\sigma} \rangle_1 = 
g_1\left(\bar{1}\bar{1}^+\right)+
\frac{1}{\langle Y^{00} \rangle_{-1}}
\left[
-\Pi_0\left(\var{1}{\sigma\sigma};\var{1}{00}\right)+
g^2_0(\bar{1}\bar{1}^+)-g_0(\bar{1}\bar{1}^+)
\right],
\label{y1}
\eeq
\beq
\Pi_0\left(\var{2}{\alpha\alpha};\var{1}{00}\right)=-
g_0(\bar{2}\bar{3})\gamma_{e,0}\left(\var{3}{\alpha}
\var{4}{\alpha};\var{1}{00}\right)
g_0(\bar{4}\bar{2}^+),
\eeq
\beq
\Pi_0\left(\var{2}{\alpha\alpha};\var{1}{\sigma\sigma}\right)=-
g_0(\bar{2}\bar{3})\gamma_{e,0}\left(\var{3}{\alpha}
\var{4}{\alpha};\var{1}{\sigma\sigma}\right)
g_0(\bar{4}\bar{2}^+),
\eeq
\beq
\Pi_0\left(\var{2}{\sigma\alpha};\var{1}{\alpha\sigma}\right)=-
g_0(\bar{2}\bar{3})\gamma_{e,0}\left(\var{3}{\alpha}
\var{4}{\sigma};\var{1}{\alpha\sigma}\right)
g_0(\bar{4}\bar{2}^+),
\eeq
\beqn
\gamma_{e,0}\left(\var{1}{\sigma}
\var{2}{\sigma};\var{3}{00}\right)&=&
-\delta(\bar{1}-\bar{2})\delta(\bar{1}-\bar{3}) +
[\delta(\bar{1}-\bar{2})t(\bar{1}-\bar{4})+
t(\bar{1}-\bar{2})\delta(\bar{1}-\bar{4})]
\nonumber\\
&&\times
g_0(\bar{4}\bar{5})
\gamma_{e,0}\left(\var{5}{\sigma}\var{6}{\sigma};\var{3}{00}\right)
g_0(\bar{6}\bar{1}^+),
\eeqn
\beqn
\gamma_{h,0}\left(\var{1}{\sigma}
\var{2}{\sigma};\var{3}{00}\right)&=&
-\delta(\bar{1}-\bar{2})\delta(\bar{1}-\bar{3}) +
g_0(\bar{2}\bar{4})
\gamma_{h,0}\left(\var{4}{\sigma}\var{5}{\sigma};\var{3}{00}\right)
g_0(\bar{5}\bar{6}^+)
\nonumber\\
&&\times
[\delta(\bar{1}-\bar{2})t(\bar{6}-\bar{2})+
t(\bar{1}-\bar{2})\delta(\bar{6}-\bar{2})],
\eeqn
\beq
\gamma_{e,0}\left(\var{1}{\sigma}
\var{2}{\alpha};\var{3}{\sigma\alpha}\right)=
\delta(\bar{1}-\bar{2})\delta(\bar{1}-\bar{3}),
\eeq
\beq
\gamma_{e,0}\left(\var{1}{\alpha}
\var{2}{\alpha};\var{3}{\sigma\sigma}\right)=0,
\eeq
\beqn
\Gamma_{1}\left(\var{1}{\sigma}
\var{2}{\sigma};\var{3}{00}\right)&=&
\frac{1}{\langle Y^{00} \rangle_{-1}}
\gamma_{e,0}\left(\var{1}{\sigma}
\var{2}{\sigma};\var{3}{00}\right)
+\frac{N}{\langle Y^{00} \rangle^2_{-1}}
\Pi_0\left(\var{1}{\sigma\sigma};\var{3}{00}\right)
g_0^{-1}(\bar{1}\bar{2})
\nonumber\\
&=&
\frac{1}{\langle Y^{00} \rangle_{-1}}
\gamma_{h,0}\left(\var{1}{\sigma}
\var{2}{\sigma};\var{3}{00}\right)
+\frac{N}{\langle Y^{00} \rangle^2_{-1}}
g_0^{-1}(\bar{1}\bar{2})
\Pi_0\left(\var{2}{\sigma\sigma};\var{3}{00}\right),
\label{Gamma1}
\eeqn
\beq
\Gamma_1\left(\var{1}{\sigma}
\var{2}{\alpha};\var{3}{\sigma\alpha}\right)=
\frac{1}{\langle Y^{00} \rangle_{-1}}
\gamma_{e,0}\left(\var{1}{\sigma}
\var{2}{\alpha};\var{3}{\sigma\alpha}\right),
\eeq
\beq
\Gamma_{1}\left(\var{1}{\alpha}
\var{2}{\alpha};\var{3}{\sigma\sigma}\right)
=+\frac{N}{\langle Y^{00} \rangle^2_{-1}}
\Pi_0\left(\var{1}{\alpha\alpha};\var{3}{\sigma\sigma}\right)
g_0^{-1}(\bar{1}\bar{2})
=0.
\label{fin1}
\eeq
Furthermore $g_1$ is related to $g_0$ by
\beq
g_1(\bar{1}\bar{2})= g_0(\bar{1}\bar{3})\Sigma_1(\bar{3}\bar{4}) 
g_0(\bar{4}\bar{2}),
\eeq
which yields a self-consistent equation 
for $g_1$, or, equivalently,
for $\Sigma_1$. 

Using Eqs.(\ref{in1}-\ref{fin1}) we obtain for
$\Sigma_1^{\sigma\sigma}$:
\beqn
\Sigma_1^{\sigma\sigma}(\bar{1}\bar{2})&=&
-\frac{[\langle Y^{00} \rangle_{0}
+\langle Y^{\sigma\sigma} \rangle_{0}]}{N}t(\bar{1}-\bar{2})
+ \delta(\bar{1}-\bar{2})t(\bar{1}-\bar{3})\left[
g_1(\bar{3}\bar{1}^+)
-\frac{g_0(\bar{3}\bar{1}^+)}{N}\right]
\nonumber\\
&&
\left.\left.
+\left[\frac{t(\bar{1}-\bar{2})}{\langle Y^{00} \rangle_{-1}}
-\frac{t(\bar{1}-\bar{3})}{N}g_0(\bar{3}\bar{4})t(\bar{4}-\bar{2})\right]
\right[
\Pi_0\left(\var{2}{\alpha\alpha};\var{1}{00}\right)-
\Pi_0\left(\var{2}{\sigma\alpha};\var{1}{\alpha\sigma}\right)
\right]
\nonumber\\
&&
+\frac{t(\bar{1}-\bar{3})}{N}g_0(\bar{3}\bar{4})
\gamma_{e,0}\left(\var{4}{\alpha}\var{5}{\alpha};\var{2}{00}\right)
g_0(\bar{5}\bar{1})g_0(\bar{1}\bar{6})t(\bar{6}-\bar{2})
\nonumber\\
&&
+\frac{t(\bar{1}-\bar{3})}{N}g_0(\bar{3}\bar{2})
g_0(\bar{2}\bar{4})
\gamma_{h,0}\left(\var{4}{\alpha}\var{5}{\alpha};\var{1}{00}\right)
g_0(\bar{5}\bar{6})t(\bar{6}-\bar{2}) \nonumber\\
&&
- \frac{t(\bar{1}-\bar{3})}{N}g_0(\bar{3}\bar{4})
h(\bar{4}\bar{5};\bar{1}\bar{2})g_0(\bar{5}\bar{6})
g_0(\bar{1}\bar{2})g_0(\bar{2}\bar{7})t(\bar{6}-\bar{2})
\label{eqeq}
\eeqn
In Eq. (\ref{eqeq}) we have used the explicit expression for the
term $\delta\Sigma^{\sigma\sigma}_1(\bar{1}\bar{2})$ calculated in
appendix B where the function $h(\bar{4}\bar{5};\bar{1}\bar{2})$
is also defined.

The expectation value of the particle number operator with spin $\sigma$
is $n_\sigma = \langle Y^{\sigma\sigma}\rangle_0 + \langle Y ^{\sigma\sigma}
\rangle_1 = n_{0\sigma}+n_{1\sigma}$, where the expectation values are
given by Eqs.(\ref{y0}) and (\ref{y1}). 
Writing 
$\gamma(\bar{1}\bar{2};\bar{3})=\gamma(\bar{1}-\bar{2},\bar{1}-\bar{3})$,
$\Pi(\bar{1}\bar{2})=\Pi(\bar{1}-\bar{2})$ and
$h(\bar{1}\bar{2};\bar{3}\bar{4}) = 
h(\bar{4}-\bar{2},\bar{4}-\bar{1},\bar{1}-\bar{3})$,
performing Fourier transforms, and adding the
contributions of $O(1)$ and $O(1/N)$ the self-energy becomes
\beqn
\Sigma^{\sigma\sigma}(k)&=&\frac{N-1}{N} \lambda_0 + \delta\lambda_1+
\left[-1+{{N-1}\over N}2n_{0\sigma}+2n_{1\sigma}\right]
\frac{t({\bf k})}{2} 
\nonumber\\
&&
+\frac{2}{N(1-2n_{0\sigma})}
\sum_{q} t({\bf k+q})\Delta\Pi(q)
-\frac{1}{N} \sum_{q} t^2({\bf k+q})g_0(k+q)\Delta\Pi(q)
\nonumber\\
&&+
\frac{1}{N}\sum_{p,q} t({\bf k-q})t({\bf p+q})g_0(k-q)g_0(p+q)g_0(p)
[\gamma_{e,0}(p,q)+\gamma_{h,0}(p+q,-q)]
\nonumber\\
&&+\frac{1}{N}\sum_{p,q}
t({\bf p}) g_0(k+q)) g_0(p+q) g_0(p) \gamma_{e,0}(p,q)\frac{c(q)}{1+b(q)},
\label{sitot}
\eeqn
where $\Delta\Pi(q)=\Pi^{\sigma\sigma}_{00}(q)-
\Pi^{\sigma\alpha}_{\sigma\alpha}(q)$ and
\beq
\delta\lambda_1 =
\sum_k t({\bf k}) g_1(k)
e^{i\omega_n 0^+} =
\sum_k t({\bf k}) g_0^2(k)
\Sigma_1(k).
\label{l1}
\eeq
Here we wrote the second index pair of $\Pi(q)$ as an subscript
to simplify the notation.
Explicit expressions for the functions $\gamma$, $\Pi$, $h$ are
\beq
\gamma_{e,0}(k,q)=
-\frac{1+b(q)-a(q)t({\bf k})}{[1+b(q)][1+b(-q)]-a(q)c(q)},
\eeq
\beq
\gamma_{h,0}(k,q)=
-\frac{1+b(-q)-a(q)t({\bf k+q})}{[1+b(q)][1+b(-q)]-a(q)c(q)},
\eeq
\beq
\Pi^{\sigma\sigma}_{00}(q)=
-\frac{a(q)}{[1+b(q)][1+b(-q)]-a(q)c(q)},
\eeq
\beq
\Pi^{\sigma\alpha}_{\sigma\alpha}(q)=
a(q),
\eeq
\beq
h(k_1,q,k_2)=\gamma_{e,0}(k_1,q)g_0(k_2)
\left[t({\bf k}_2)-\frac{c(q)}{1+b(q)}\right],
\eeq
and for the susceptibilities $a$, $b$, $c$\cite{Kulic1} 
\beq
a(q)=-\sum_k g_0(k)g_0(k+q),
\eeq
\beq
b(q)=-\sum_k t({\bf k}) g_0(k)g_0(k+q),
\eeq
\beq
c(q)=-\sum_k t({\bf k}) t({\bf k+q}) g_0(k)g_0(k+q).
\eeq
Note that $\gamma_{e,0}(p,q)=\gamma_{h,0}(p+q,-q)$.

Finally, the relation between the particle number $n_\sigma$ and 
the chemical potential becomes, using Eq.(\ref{y1}),
\beq
n_\sigma = n_{0\sigma}\Bigl(1-{{2-2n_{0\sigma}}\over{N(1-2n_{0\sigma})}}\Bigr)
-{2\over{N(1-2n_{0\sigma})}}\sum_q\Pi^{\sigma\sigma}_{00}(q)+\delta 
n_{1\sigma},
\label{ntot}
\eeq
where $\delta n_{1\sigma}$ is defined by
\beq
\delta n_{1\sigma} = \sum_k g_1(k) 
e^{i\omega_n 0^+} =
\sum_k g_0^2(k) \Sigma_1^{\sigma\sigma}(k).
\label{deltan1}
\eeq

Eq.(\ref{sitot}) gives an explicit expression for the self-energy 
taking into account the orders $O(1)$ and $O(1/N)$ of the $1/N$ expansion.
The first two terms represent renormalizations of the chemical potential.
The third and fourth terms are frequency-independent 
and thus contribute only to the
dispersion of quasi-particles. Terms 5-7 are
frequency-dependent,
contribute to the damping of quasi-particles, and cannot simply be obtained
by a factorization or mode-coupling assumption applied to Eq.(\ref{seinvert}).

Eq.(\ref{ntot}) represents the relation between the particle number
and the chemical potential. As can be seen from Eqs.(\ref{y0}) and (\ref{y1})
$n_\sigma$ is not only determined by the zeroth and first orders
of the normalized Green's function $g$ but by additional terms.
The appearance of such
additional terms is not unexpected because $n_\sigma$
is given by the normalized Green's function $g({\bar 1}{\bar 1}^+)$
at $N=\infty$ and by the Green's function $G({\bar 1}{\bar 1}^+)$ at $N=2$
so that the $1/N$ expansion must interpolate between these two cases.

\section{Violation of Luttinger's theorem in O(1/N)}

Eq.(\ref{sitot}) shows that the imaginary part of the self-energy
vanishes at the chemical potential. There is thus a Fermi surface $\{
{\bf k}_F\}$ determined by the condition
\beq
\mu = \Sigma({\bf k}_F,\omega=0,\mu),
\label{fermi}
\eeq
where the chemical potential $\mu$ is determined by the particle
number $n = N n_\sigma$ per site. $n_\sigma$ does not depend 
on the spin direction $\sigma$ in the normal state, of
course. Nevertheless, it is convenient to denote the particle number
per spin direction by $n_\sigma$ to distinguish it from the total
particle number $n$. In Eq.(\ref{fermi}) we have written 
explicitly the dependence of $\Sigma$ on the wave vector $\bf k$,
the frequency $\omega$ and the chemical potential $\mu$
and omitted unnecessary spin labels. 

In the unconstrained, free case the self-energy $\Sigma$
in Eq.(\ref{fermi}) should be replaced by $-t({\bf k})/N$.
Each state $\bf k$ can be occupied by $N$ particles. 
The set of Hamiltonians with different $N's$ used for the $1/N$
expansion should correspond to the same $n_\sigma$.
As a result the free Fermi surface $\{{\bf k}_F\}$ is independent of $N$
though the corresponding chemical potential $\mu$ and the electron dispersion
depend on $N$. Turning on the interaction between electrons
the Fermi surface depends in general on $N$. Luttinger's theorem
states, however, that the volume enclosed by the Fermi surface is
independent of $N$ and also of the interaction strength.

In the constrained case one expands for a fixed $n_\sigma$ the quantities 
$\mu$, $\Sigma$, and ${\bf k}_F$ in powers of $1/N$ similar as in
Eqs.(\ref{expansion1})-(\ref{expansion2}). Using Eqs.(\ref{s0k})
and (\ref{y00}) the zeroth order of Eq.(\ref{fermi}) is
\beq
\mu_0 = -({1\over2}-n_{0\sigma})t({\bf k}_{0F}) +\lambda_0(\mu_0).
\label{fermi0}
\eeq
Eqs.(\ref{g0}) and (\ref{y0}) yield at $T=0$
\beq
n_{0\sigma} = \sum_{\bf k} \Theta(\mu_0 -\lambda_0 +({1 \over 2}-n_{0\sigma})
t({\bf k})).
\label{n0}
\eeq
For a fixed $n_{0\sigma}$ Eq.(\ref{n0}) determines $\mu_0-\lambda_0$
and Eq.(\ref{fermi0}) ${\bf k}_{0F}$. These equations are the same as
in the unconstrained, free case if we choose 
$({1\over2}-n_{0\sigma})t({\bf k})$ for the dispersion and renormalize
the chemical potential appropriately. This means that at $N=\infty$ the Fermi 
surface is
identical to that of the free, unconstrained case and that, in
particular, Luttinger's theorem holds. This result is not very
surprising because at $N=\infty$ the particles are renormalized but do not 
interact with each other.

The first-order of Eq.(\ref{fermi}) in $1/N$ is given by
\beq
\mu_1 = {{\partial \Sigma_0({\bf k}_{0F},\mu_0)} \over {\partial \mu_0}}
\cdot \mu_1 + {{\partial \Sigma_0({\bf k}_{0F},\mu_0)} \over 
{\partial {\bf k}_{0F}}} \cdot {\bf k}_{1F} +
\Sigma_1({\bf k}_{0F},\omega=0,\mu_0).
\label{fermi1}
\eeq
Here ${\bf k}_{1F}$ is assumed to be parallel to the vector
${\partial \Sigma_0}/{\partial{\bf k}_{0F}}$ with ${\bf k}_{0F}$ as origin.
Because $\Sigma_0$ is independent of frequency we have dropped
the frequency argument in $\Sigma_0$ for convenience. The
change $\mu_1$ in the chemical potential is to be calculated for
a fixed $n_\sigma$, i.e., from
\beq
(\partial n_{0\sigma}/\partial \mu_0)\cdot \mu_1 + n_{1\sigma}(\mu_0) = 0.
\label{mu1}
\eeq
Eq.(\ref{fermi1}) must hold for any point ${\bf k}_{0F}$
on the Fermi surface. This equation thus determines ${\bf k}_{1F}$
as a function of ${\bf k}_{0F}$ and thus the change of the Fermi
surface in $O(1/N)$. Note that due to the last term in Eq.(\ref{fermi1})
the shape of the Fermi surface will in general change by the
interaction between the particles. 

Eq.(\ref{fermi1}) can be simplified in the following way. $\Sigma$
depends on $\mu$ only via $\mu-\lambda$ where $\lambda$ is again a function
of $\mu$. Defining a renormalized chemical potential $\tilde{\mu}$
by
\beq
\tilde{\mu} = \mu -\lambda,
\eeq
we can drop $\lambda$ in $\Sigma$ everywhere replacing $\mu$ by
$\tilde{\mu}$. 
Eq.(\ref{mu1}) also holds if $\mu_0$ and $\mu_1$ are replaced by
$\tilde{\mu_0}$ and $\tilde{\mu_1}$, respectively, yielding
\beq
(\partial n_{0\sigma}/\partial \tilde{\mu}_0)\cdot \tilde{\mu}_1 + 
n_{1\sigma}(\tilde{\mu}_0) = 0.
\label{mu2}
\eeq
It is then easy to see
that the first term in Eq.(\ref{fermi1}), written in terms of $\tilde{\mu}$,
cancels the term $n_{1\sigma}
t({\bf k})$ in Eq.(\ref{sitot}). Eq.(\ref{fermi1}) thus becomes
\beq
\tilde{\mu}_1 = {{\partial \Sigma_0({\bf k}_{0F},\tilde{\mu_0})} \over 
{\partial {\bf k}_{0F}}} \cdot {\bf k}_{1F} +
\tilde{\Sigma}_1({\bf k}_{0F},\omega=0,\tilde{\mu}_0),
\label{fermi2}
\eeq
where $\tilde{\Sigma}_1$ is given by Eq.(\ref{sitot}) with all terms in
the first line on the right-hand side dropped except for the term
$-n_{0\sigma}t({\bf k})/N$.
According to Eq.(\ref{n0}) the derivative $\partial n_{0\sigma}/
\partial \tilde{\mu_0}$ gives rise to two contributions, one in which
the derivative acts on $\tilde{\mu}_0$ and one, where it acts on 
$n_{0\sigma}$. We also split
$n_{1\sigma}$, given by Eq.(\ref{ntot}), into three contributions,
\beq
n_{1\sigma} = \tilde{n}_{1\sigma} + \delta \tilde{n_{1\sigma}} +
\delta \bar{n}_{1\sigma},
\label{n1}
\eeq
with
\beq
\tilde{n}_1 = -{{4n_{0\sigma}(1-n_{0\sigma})}\over
{N(1-2n_{0\sigma})}}(1+\sum_q\Pi^{\sigma\sigma}_{00}(q)),
\label{nn1}
\eeq
\beq
\delta \tilde{n}_{1\sigma} = 
\sum_k g_0^2(k) \tilde{\Sigma}_1^{\sigma\sigma}(k),
\label{deltan2}
\eeq
\beq
\delta {\bar n}_{1\sigma} = 
\sum_k g_0^2(k) n_{1\sigma} t({\bf k}).
\label{deltan3}
\eeq
Using these results Eq.(\ref{mu1}) can be rewritten as
\beq
{{\partial n_{0\sigma}}\over{\partial \tilde{\mu_0}}}\mid_{ex} \cdot
\tilde{\mu_1} + \sum_k g_0^2(k) \Bigl({{\partial n_{0\sigma}}\over{\partial
\tilde{\mu_0}}}\Bigr) t({\bf k}) \tilde{\mu}_1 + \tilde{n}_{1\sigma}
+\delta \tilde{n}_{1\sigma} + \delta {\bar n}_{1\sigma} = 0,
\label{mu3}
\eeq
where $\mid_{ex}$ means that the derivative should be taken only with
respect to the explicit dependence on ${\tilde \mu}_0$. 
>From Eq.(\ref{mu2}) and (\ref{deltan3}) follows that the second and fifth 
terms on the left-hand side of Eq.(\ref{mu3}) cancel each other. Solving
Eq.(\ref{deltan3}) for $\tilde {\mu_1}$ we insert its solution
into Eq.(\ref{fermi2}) and average the resulting equation over the
zeroth-order Fermi surface and obtain
\beq
-\tilde{n}_{1\sigma} - \delta \tilde{n}_{1\sigma} = 
\sum_{\bf k} \delta(\epsilon_{\bf k}-\tilde{\mu}_0)
{{\partial \epsilon_{\bf k}}\over{\partial {\bf k}}} \cdot {\bf k}_{1F}
+\sum_{\bf k} \delta(\epsilon_{\bf k}-\tilde{\mu}_0) \tilde{\Sigma}_1
({\bf k},\omega = 0,\tilde{\mu}_0),
\label{fermi3}
\eeq
where $\epsilon_{\bf k}$ denote the zeroth-order one-particle 
energies. The first term on the right-hand side of Eq.(\ref{fermi3}) 
is equal to the Fermi surface
integral $\int d{\bf S_F} \cdot {\bf k}_{1F} = V_1^F$, where ${\bf S_F}$
is a directed element of the Fermi surface and $V_1^F$ is the 
change in $O(1/N)$ of the volume enclosed by the Fermi surface. 
Eq.(\ref{fermi3}) becomes finally
\beq
V_1^F = -\tilde{n}_{1\sigma} - \sum_{k} g^2_0(k) \Big( 
\tilde{\Sigma}_1(k,\tilde{\mu_0}) 
-\tilde{\Sigma}_1({\bf k},\omega= 0,\tilde{\mu}_0)
\Big).
\label{V1}
\eeq

The first term on the right-hand side of Eq.(\ref{V1}) is due to the
unusual relation between the particle occupation $n_\sigma$ and the
chemical potential $\tilde{\mu}$ specified in Eq.(\ref{ntot}). In Fermi
liquid theory $n_\sigma$ is given by the one-particle Green's function
integrated over all momenta and over frequency up to $\tilde{\mu}$. 
Such a relation holds in our case only at $N=\infty$ and $N=2$
with $g$ and $G$ to be used as Green's function, respecively.
The $1/N$ expansion must interpolate between these two cases
which is the origin of the more involved nature of Eq.(\ref{ntot}). Thus the
first term in Eq.(\ref{V1}) is intimately linked to the $1/N$ expansion
and of fundamental nature. In Fermi liquid theory it is, of course, absent.

In the second term on the right-hand side of Eq.(\ref{V1}) any
frequency-independent contribution to $\tilde{\Sigma}_1$ drops
out. This term has the same form as
the term in Fermi liquid theory where the product of the frequency
derivative of the self-energy and the Green's function is integrated
over (second term in the square bracket in Eq.(19.12) in
Ref.\cite{Abrikosov}). The only difference is that in our case the 
involved Green's function is the zeroth-order Green's function $g_0$
and $\Sigma$ is a functional of $g_0$. This term vanishes in Fermi liquids for
general reasons yielding Luttinger's theorem. Though we are presently
unable to prove something similar for the second term in Eq.(\ref{V1})
it is easy to see that it cannot cancel in general the first term on 
the right-hand side of Eq.(\ref{V1}). 
>From Eq.(\ref{nn1}) follows that $\tilde{n}_1$
diverges in the limit $n_{0\sigma} \rightarrow 1/2$. In contrast to that
the second term in Eq.(\ref{V1}) is clearly regular in that limit.
>From this one
concludes that there exists at least a finite, $N$-independent interval 
in doping
near half-filling where the first term in Eq.(\ref{V1}) dominates
and makes $V_1^F$ non-zero. Assume now that $V_1^F$ is zero in an 
arbitrarily small interval of the doping $\delta =1 -2n_{0\sigma}$. 
This implies that the regular
part of $V_1^F$ is equal to $A/\delta$ with some constant $A$ throughout
this interval. Performing an analytic continuation in the variable $\delta$
in $\Sigma_1$ and $V_1^F$ shows that the regular part should be
given by $A/\delta$ in the whole interval $0 < \delta < 1$ if there is
no phase transition present making the interchange of sums over $k$ and the   
continuation in $\delta$ impossible. The regular part in $V_1^F$ would
then develop a singularity at $\delta = 0$ in contradiction to our
assumptions. 
Since detailed calculations show\cite{Zeyher3,Cappelluti} that at $U = \infty$
and large $N$'s there are no phase transitions one concludes
that the Luttinger theorem must be violated for dopings throughout the
entire interval $0 < \delta < 1$. 

\begin{figure}[t]
\centerline{\psfig{figure=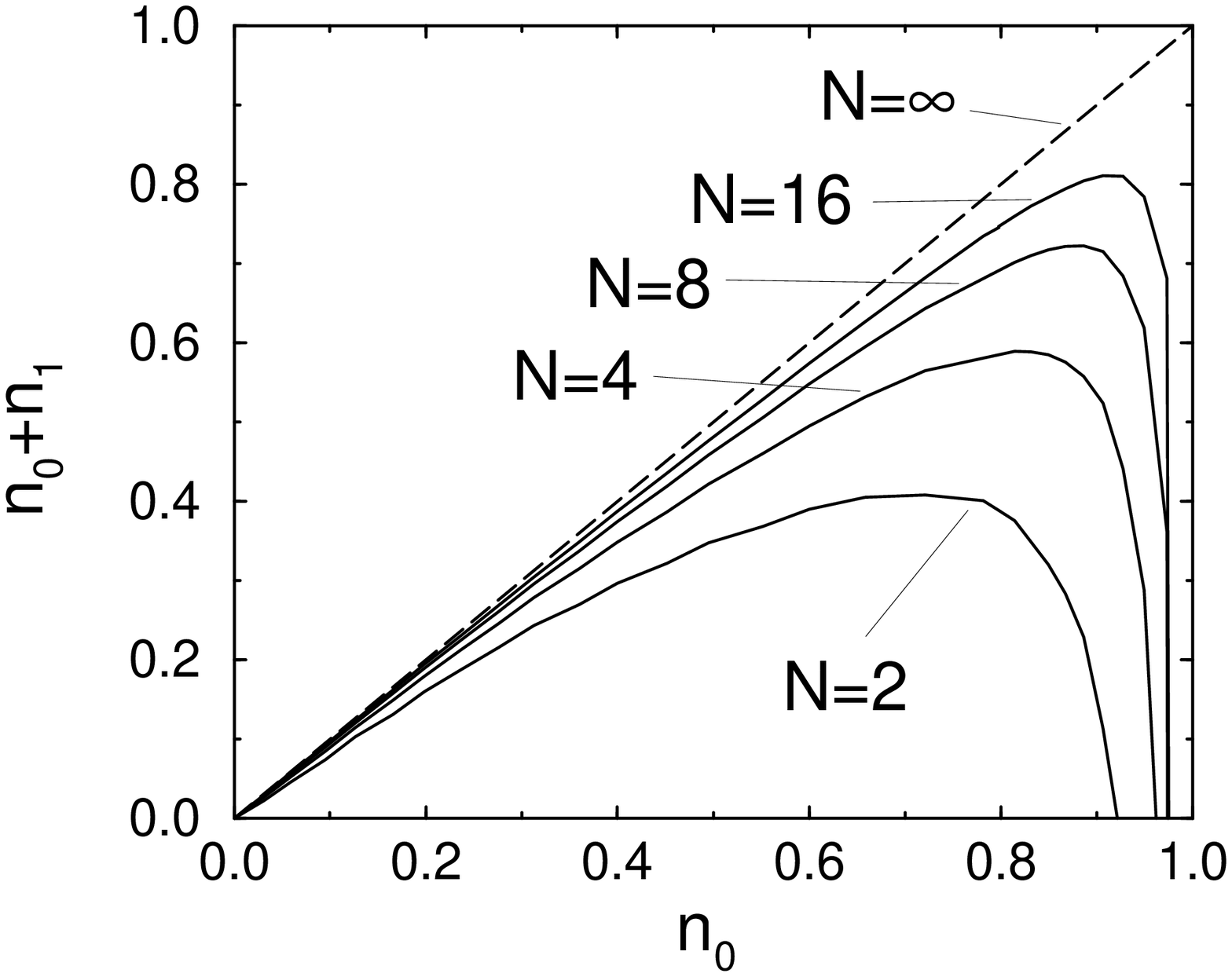,width=7.6cm}}
\caption{Sum of the $O(1)$ and $O(1/N)$ contributions $n_0$ and
$n_1$, respectively, to the site occupancy  as a function of $n_0$ for
different $N$'s. Only terms which diverge near half-filling haven been
considered in $n_1$.}
\end{figure}

Fig. 1 shows the dependence of the sum of $n_{0}$ and 
$n_1$ as a function of $n_{0}$ for $N=\infty$ (dashed line) and
$N=16,8,4,$ and $2$ (solid lines). $n_1$ is always negative which means
that for a given $n_0$ or $\tilde{\mu}_0$ less particles can be accomodated
at a site if electronic correlations are taken into account compared to
the uncorrelated case. This effect is 
very small at small $n_0$'s but increases monotonously towards half-filling
where $n_1$ diverges. For $N=2$ a maximal site occupation of only about 0.4 is
possible. For larger $\tilde{\mu}_0$ the $n(\tilde{\mu}_0)$ relation    
has a maximum and then a negative slope signalizing an unstable 
homogenous state.
This result clearly shows that a calculation which includes only 
$O(1)$ and $O(1/N)$ contributions cannot be used to describe
the physical case of $N=2$ near half-filling. With increasing $N$
the maximal site occupancy increases monotonously towards 1 and the
maximum of the curves is shifted towards $n_0=1$.

Fig. 2 shows the total volume $V^F$ enclosed by the Fermi surface and 
calculated in $O(1)$ and $O(1/N)$ keeping the singular first
term in Eq.(\ref{V1}). The dashed line corresponds to
$N=\infty$ and represents exactly the volume predicted by the Luttinger
theorem. The solid lines describe the cases $N=16,8,4,$ and $2$.
The Figure shows that electronic correlations, taken into account in
$O(1/N)$, always increase $V^F$. This increase is small
\begin{figure}[b]
\centerline{\psfig{figure=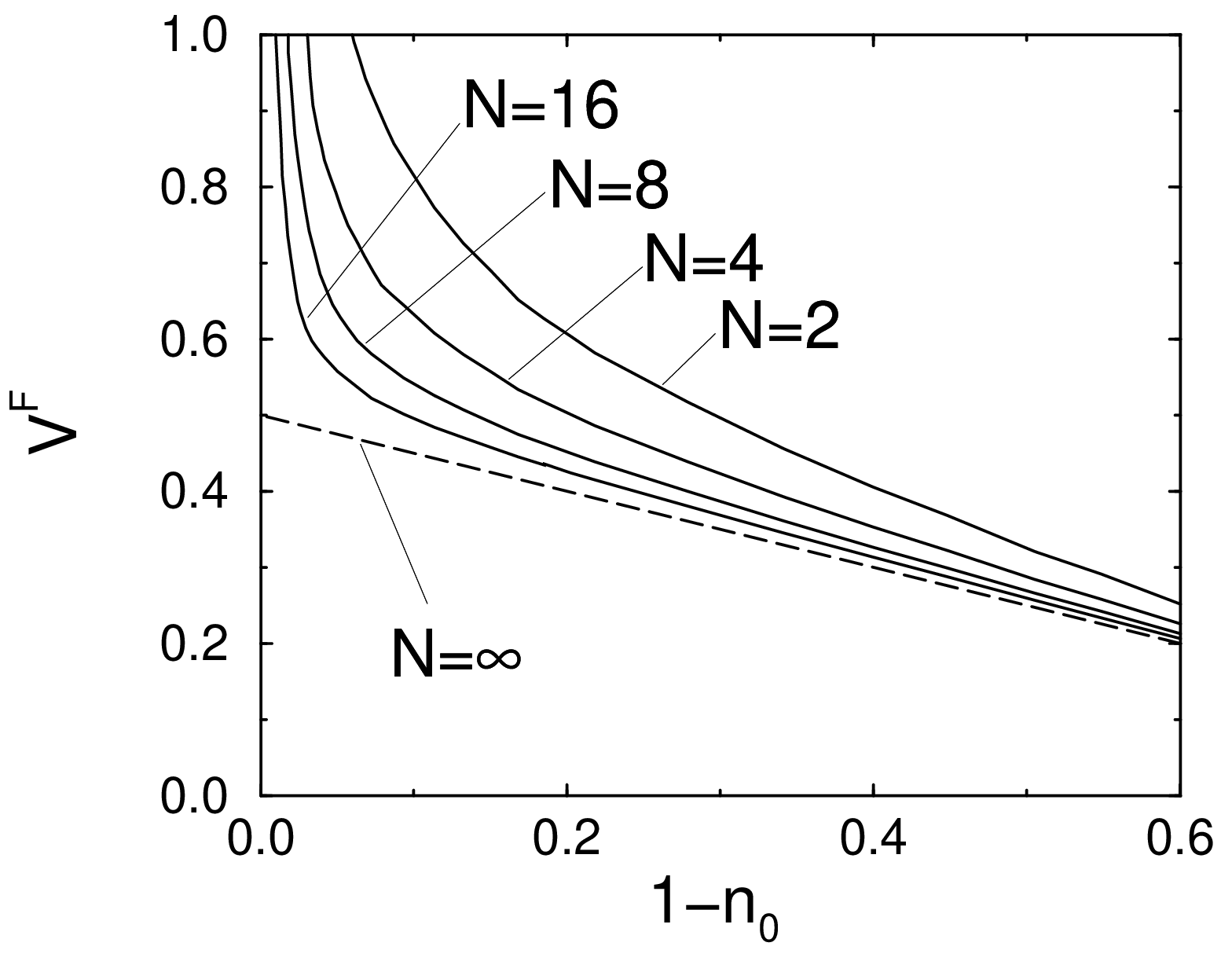,width=7.6cm}}
\caption{Sum of the $O(1)$ and $O(1/N)$ contributions to the Luttinger
volume $V^F$ as a function of the doping $\delta = 1 - n_0$ keeping
only divergent terms in the $O(1/N)$ contribution. The dashed line 
corresponds to $N=\infty$ and obeys Luttinger's theorem.}
\end{figure}
at small electron concentrations but becomes larger and larger at small
dopings. Our results are asymptotically exact at sufficiently large $N$'s
and in a finite, $N$-independent doping interval near half-filling.
As argued above this implies that Luttinger's theorem is violated throughout
the entire doping interval $0 < \delta < 1$. This means that the solid
curves for large $N$'s are never coincident with the dashed line
in Fig. 2 and that these lines are at least qualitatively a good
approximation for all dopings. Fig. 2 illustrates also the fact that
increasing electronic correlations increases the Luttinger volume $V^F$,
i.e., $V_1^F$ is positive. The solid lines in Fig. 2 diverge towards
half-filling due to the $1/N$ expansion of the reciprocal spectral
weight $Q^{-1}$. The convergence radius of this expansion
is determined
by $1/N < \delta/n_0$ and thus becomes very small at small dopings. 
This means that for a fixed $N$ the power expansion in $1/N$ breaks down
at small dopings. The larger $N$ is the smaller is the critical value for
the doping where this expansion breaks down.  

\section{Conclusion}

Using a $1/N$ expansion we have derived explicit expressions for the
$O(1)$ and $O(1/N)$ contributions, $\Sigma_0$ and $\Sigma_1$, respectively,
for the self-energy of a $SU(N)$ symmetric Hubbard model with $N$
degrees of freedom and $U=\infty$. Using previously derived rules these 
results can
immediately be generalized to the case of a finite, but large $U$. We find
that the frequency-dependent part of $\Sigma$ is regular in the doping
$\delta$,
whereas the frequency-independent part develops a singularity in $\delta$
due to the expansion of the inverse spectral weight $Q^{-1}$ 
in powers of $1/N$. This means that near half-filling large $N$'s are
required to obtain quantitative results. As an example we find
that the $n(\mu)$ curve, calculated with $\Sigma_0$ and $\Sigma_1$ and
$N=2$, becomes unreliable already near and below a doping $\delta \sim\
0.20$. Our derivation of $\Sigma$ also shows that $\Sigma_1$ involves
rather sophisticated vertex corrections and that usual many-body
assumptions (neglect of vertex corrections, lowest-order iterations
of the functional equations for the self-energy and the vertex)
cannot be used in order not to miss cancellation effects. These
features are connected to the fact that the hopping $t$ cannot be considered
as a small parameter in the perturbation expansion in the kinetic
energy of the electrons, at least in the metallic state. This is related
to the fact that $t$ appears not only in the numerator but also in the
denominator of the perturbation series because the dispersion of the 
one-particle energies is also determined by $t$.
 
Using the above results we have studied the $1/N$ expansion for the
Luttinger volume $V^F = V_0^F + V_1^F + ...$ where $V^F$ denotes the
volume enclosed by the Fermi surface. The $O(1)$ contribution $V_0^F$
obeys Luttinger's theorem but this theorem is violated already in $O(1/N)$.
We have shown that $V_1^F \neq 0$ throughout the entire doping interval
$0 < \delta < 1$, if the normal state is stable, and that $V_1^F > 0$ 
at least  in a finite, $N$-independent interval near half-filling. 
Keeping only terms which are leading at small dopings our calculations
suggest that $V_1^F > 0$ for all dopings and that $V_1^F$ becomes very
small at large dopings. There are also good reasons to believe that
the normal state to which our discussions is restricted, can be 
considered as the stable state at large $N$'s throughout the interval
$0 < \delta < 1$. The largest transition temperatures to superconducting 
states behave roughly as $\sim U^{-2}$ at large $U$'s\cite{Zeyher3}. 
Strictly speaking, this is not entirely true because extremely 
weak superconducting instabilities have been found even for $U=\infty$
\cite{Zeyher4}. The corresponding transition temperatures are, however,
astronomically small so that these instabilities can be neglected 
for our purposes.  In a similar way all the other instabilities
\cite{Cappelluti} of the $t-J$
model at large $N$'s such as flux phases, bond-order or charge density waves,
vanish in the limit $ J  \rightarrow 0$ or $U \rightarrow \infty$.
This means that we have indeed dealt with the most stable ground state of
our system.  

Finally we want to point out that our results agree with two recent
numerical studies of the Luttinger volume in the Hubbard model. In Ref.
\cite{Eder} Quantum Monte-Carlo simulations for the 2D Hubbard model with 
$U/t=8$ and $T/t=0.33$ have been preformed and an increase in $V_F$
beyond the value predicted by the Luttinger theorem has been found
for $\delta \leq 0.2$. Since at these high temperatures structures due
to the quasi-particle peak and the incoherent background cannot be
distinguished the interpretation of the results is not unique: a) $V_F$
may be the volume enclosed by the true Fermi surface or b) $V_F$
is a volume enclosing a ``spectral weight Fermi surface'' and thus be
related to the spectral distribution of the whole Green's function.
Similar results, subject to a similar uncertainty in the interpretation,
have been found in Ref.\cite{Putikka} using high-temperature expansions
and considering the momentum distribution function. In our case it is clear 
that only
the case a) can apply and that $V_F$ denotes the volume enclosed by a
constant energy contour of quasi-particles.

\newpage
\begin{appendix}
\section{Cancellation of reducible self-energy contributions}

In Eq. (\ref{dersp}) we have omitted the following reducible
self-energy contributions:
\[
\Sigma^{red}_e(12) =
-\int d1'd2'd3d3'd4 t_e(132')\langle Y(3)\rangle G(2'1')
\langle Y(3')\rangle t_h(43'1')Q^{-1}(42)\]
\[+\int d1'd2'd3d3'd4'd5'd6 t_e(132')\langle Y(3)\rangle G(2'1') 
\Gamma(1'4';3') G(4'5')
t_h(63'5')Q^{-1}(62)\]
\[+\int d1'd2'd3d3'd4d5d6 t_e(134) G(45) \Gamma(52';3) G(2'1')
\langle Y(3')\rangle t_h(63'1')Q^{-1}(62)\]
\[-\int d1'd2'd3d3'd4d4'd5d5'd6 t_e(134)G(45)\Gamma(52';3)G(2'1')
\Gamma(1'4';3')G(4'5') t_h(63'5')Q^{-1}(62)\]
\[-\int d1'd2'd3d4d5d6 t_e(134) G(45) \Gamma(52';3) G(2'6)Q^{-1}(61')
\Sigma_e(1'2)\]
\beq
+\int d1'd2'd3d4 t_e(132')\langle Y(3)\rangle G(2'4)Q^{-1}(41')
\Sigma_e(1'2).
\label{reduc}
\eeq
In equilibrium the factors $Q^{-1}(12)$ yield a delta function
$\delta({\bar 1}-{\bar 2})$ and a constant factor 
which can be taken outside of the integrals. Writing the expectatin value of 
three Hubbard operators in Eq.(\ref{seinvert}) in terms of a functional 
derivative
and using Eq.(\ref{Gamma}) we obtain from Eq.(\ref{seinvert})
\beq
\Sigma_e(12)= -\int d3 t_e(132) \langle Y(3)\rangle
+ \int d3d4d5 t_e(134) G(45) \Gamma(52;3).
\label{self}
\eeq
An alternative procedure to calculate $G$ uses the equation of motion
Eq.(\ref{hmot}) instead of Eq.(\ref{emot}). Eqs.(\ref{dysinvert}) is then 
replaced by the equivalent equation
\beq
G^{-1}(12) = \int d3 [G_0^{-1}(13) -\mu_h(13) + \Sigma_h(13)]Q^{-1}(32),
\label{dysonr}
\eeq
and Eq.(\ref{self}) by
\beq
\Sigma_h(12)= -\int d3 t_h(231) \langle Y(3)\rangle
+ \int d3d4d5 \Gamma(14;3) G(45) t_h(235).
\label{selfh}
\eeq
In view of Eqs.(\ref{mue}) and (\ref{muh}) Eqs.(\ref{dysinvert}) and 
(\ref{selfh})
yield $\Sigma_h(12)= - \Sigma_e(12)$. Using this identity as well as
Eqs.(\ref{self}) and (\ref{selfh}) Eq.(\ref{reduc}) can be written as
\beq
\Sigma_e^{red}(12)= - Q^{-1}\int d1'd2' \Sigma_e(12') G(2'1') \Sigma_h(1'2)
- Q^{-1}\int d1'd2' \Sigma_e(12') G(2'1') \Sigma_e(1'2) = 0.
\eeq
We thus have shown that the reducible self-energy terms cancel
each other and do not contribute to the self-energy.

\newpage
\section{Evaluation of $\delta \Sigma$ }

Let us define a function $F$ by
\beq
F(12,34)=\frac{\delta^2 G^{-1}(12)}{\delta K(3) \delta K(4)}=
\frac{\delta \Gamma(12;3)}{\delta K(4)}=
\frac{\delta \Gamma(12;4)}{\delta K(3)}.
\eeq
The equilibirum contribution $\delta \Sigma$ can then be written as
\beqn
\delta \Sigma^{\sigma\sigma}({\bar 1}{\bar 2}) &=&
-\int d{\bar 3}d{\bar 4}d{\bar 5}d{\bar 6}
\frac{t(\bar{1}-\bar{3})}{N}g(\bar{3}\bar{4})g(\bar{5}\bar{6})
\frac{t(\bar{6}-\bar{2})}{N}\times
[\langle Y^{00}\rangle + \langle Y^{\sigma \sigma}\rangle]
\nonumber \\
&&\left[
F\left(\var{4}{\sigma}\var{5}{\sigma};
\var{2}{00}\var{1}{00}\right)+
F\left(\var{4}{\sigma}\var{5}{\sigma_1};
\var{2}{\sigma\sigma_1}\var{1}{00}\right)\right.
+\left.
F\left(\var{4}{\sigma_1}\var{5}{\sigma};
\var{2}{00}\var{1}{\sigma_1\sigma}\right)+
F\left(\var{4}{\sigma_1}\var{5}{\sigma_2};
\var{2}{\sigma\sigma_2}\var{1}{\sigma_1\sigma}\right)
\right],
\label{sigf}
\eeqn
where we have written out the spin labels and summation over indices 
which do not appear on the left-hand side of this equation is assumed.
It is also convenient to introduce the function
\beq
f(12,34)=\frac{\delta^2 g_e^{-1}(12)}{\delta K(3) \delta K(4)}.
\eeq
F can then be expressed as
\beqn
F(12;34) &=&
Q^{-1}(1)f(12;34)+\frac{\delta Q^{-1}(1)}{\delta K(3)}\gamma_e(12;4)
\nonumber\\
&&
+\frac{\delta Q^{-1}(1)}{\delta K(4)}\gamma_e(12;3)
+\frac{\delta^2 Q^{-1}(1)}{\delta K(3) \delta K(4)}g_e^{-1}(12).
\eeqn

Next we expand the functions $F$, $f$ in powers of $1/N$ similar as in 
Eqs.(\ref{expansion1})-(\ref{expansion2}). Considering the $O(1/N)$
contribution of $\delta \Sigma$, i.e., $\delta \Sigma_1$, the
first term in the square bracket in Eq.(\ref{sigf}) drops out and we
obtain
\beqn
\delta \Sigma_1^{\sigma\sigma}({\bar 1}{\bar 2}) &=&
-\int d{\bar 3}d{\bar 4}d{\bar 5}d{\bar 6}
\frac{t(\bar{1}-\bar{3})}{N}g_0(\bar{3}\bar{4})g_0(\bar{5}\bar{6})
t(\bar{6}-\bar{2}){\langle Y^{00}\rangle}_{-1} \times 
\nonumber \\
&&\left[
F_1\left(\var{4}{\sigma}\var{5}{\alpha};
\var{2}{\sigma\alpha}\var{1}{00}\right)\right.
+\left.
F_1\left(\var{4}{\alpha}\var{5}{\sigma};
\var{2}{00}\var{1}{\alpha\sigma}\right)+
{1 \over{\langle Y^{00}\rangle}_{-1}}
f\left(\var{4}{\sigma_1}\var{5}{\alpha};
\var{1}{\sigma_1\sigma}\var{2}{\sigma\alpha}\right)
\right].
\label{sigf2}
\eeqn
In the last term in the square bracket in Eq.(\ref{sigf2}) 
the $O(1)$ contribution is to be taken and there is 
a sum over $\sigma_1$ whereas $\alpha$ is a fixed generic index 
$\alpha \neq \sigma$. 
Straightforward considerations show that
\beq
f_0\left(\var{1}{\sigma}\var{2}{\alpha};
\var{3}{\sigma\alpha}\var{4}{00}\right)=
f_0\left(\var{1}{\alpha}\var{2}{\sigma};
\var{3}{00}\var{4}{\alpha\sigma}\right)=0,
\label{f}
\eeq
\beq
F_1\left(\var{1}{\sigma}\var{2}{\alpha};
\var{3}{\sigma\alpha}\var{4}{00}\right) =
F_1\left(\var{1}{\sigma}\var{2}{\alpha};
\var{4}{00}\var{3}{\sigma\alpha}\right)=
\frac{N}{\langle Y^{00} \rangle^2_{-1}}
\Pi_0\left(\var{3}{\sigma\sigma};\var{4}{00}\right)
\delta(\bar{1}-\bar{2})\delta(\bar{1}-\bar{3}).
\label{F1}
\eeq

For example, consider Eq.(\ref{f}). We rewrite the functional derivative
$\delta / \delta K({{00}\atop{\bar 4}})$ as a derivative
with respect to $g$ and obtain in leading order
\beq
f_0\left(\var{1}{\sigma}\var{2}{\alpha};
\var{3}{\sigma\alpha}\var{4}{00}\right)
= {{\delta \gamma\left(\var{1}{\sigma}\var{2}{\alpha};
\var{3}{\sigma\alpha}\right)}\over{\delta 
K\left(\var{4}{00}\right)}}
= - \sum_{\beta}\int d{\bar 6}d{\bar 7} 
{{\delta \gamma\left(\var{1}{\sigma}\var{2}{\alpha};
\var{3}{\sigma\alpha}\right)}\over{\delta g\left(\var{6}{\beta}
\var{7}{\beta}\right)}} \cdot g_0({\bar 6}{\bar 4}) 
g_0({\bar 4}{\bar 7}),
\label{ff}
\eeq
where all quantities are to be taken in equilibrium once the derivatives
have been carried out. The $O(1)$ term of $\gamma$ is independent of $g$
and thus drops out. Since in the generic case $\beta \neq \sigma$,
$\beta \neq \alpha$, one internal sum in the $O(1/N)$ contribution
for $\gamma$ is removed by the derivative $\delta/\delta g\left(\var{6}
{\beta}\var{7}{\beta}\right)$
but the sum over $\beta$ gives a factor $N$. This means that the right-hand 
side of Eq.(\ref{ff}) is of order $O(1/N)$ and yields Eq.(\ref{f}).

Finally, the last term in the square bracket of Eq.(\ref{sigf2})
can be evaluated as follows. First we rewrite it in a similar way as $f_0$
in Eq.(\ref{ff}) and obtain 
\beq
f\left(\var{4}{\sigma_1}\var{5}{\alpha};
\var{1}{\sigma_1\sigma}\var{2}{\sigma\alpha}\right)
= {{\delta \gamma\left(\var{4}{\sigma_1}\var{5}{\alpha};
\var{1}{\sigma_1\sigma}\right)}\over{\delta 
K\left(\var{2}{\sigma\alpha}\right)}}
= \int d{\bar 6}d{\bar 7} 
{{\delta \gamma\left(\var{4}{\sigma_1}\var{5}{\alpha};
\var{1}{\sigma_1\sigma}\right)}\over{\delta g\left(\var{6}{\sigma}
\var{7}{\alpha}\right)}} \cdot g_0({\bar 6}{\bar 2}) 
g_0({\bar 2}{\bar 7}).
\label{gammaderiv}
\eeq
The quantity $\delta \gamma/\delta g$ in Eq.(\ref{gammaderiv})
is very similar to the quantity $g^{(2)}$ in Eq.(34) of Ref.\cite{Zeyher3}
and can also be calculated in a similar way. One obtains
\beqn 
{\Large \Sigma_{\sigma_1}} {{\delta \gamma\left(\var{4}
{\sigma_1}\var{5}{\alpha};
\var{1}{\sigma_1\sigma}\right)}\over{\delta g\left(\var{6}{\sigma}
\var{7}{\alpha}\right)}} = \delta({\bar 6}-{\bar 1}) \cdot
h({\bar 4}{\bar 5};{\bar 1}{\bar 7}),
\label{gammaderiv1}
\eeqn
where $h$ satisfies the equation
\beqn
h({\bar 4}{\bar 5};{\bar 1}{\bar 7}) &=& 
\int d{\bar 3} t({\bar 4}-{\bar3}) \gamma_{e,0}({\bar 7}{\bar 5};{\bar 4}) 
g_0({\bar 3}-{\bar 1}) \nonumber \\
&+&\int d{\bar 3}d{\bar 6} t({\bar 4}-{\bar 3}) g_0({\bar 3}-{\bar 6})
g_0({\bar 4}-{\bar 1}) h({\bar 6}{\bar 5};{\bar 4}{\bar 7}).
\label{hequation}
\eeqn
Writing $h({\bar 4}{\bar 5};{\bar 1}{\bar 7})=
h({\bar 7}-{\bar 5},{\bar 7}-{\bar 4},{\bar 4}-{\bar 1})$, and 
performing Fourier transformations, Eq. (\ref{hequation}) becomes
\beq
h(k_1,q,k_2)=t({\bf k}_2)g_0(k_2)\gamma_{e,0}(k_1,q)+
g_0(k_2)\sum_p t({\bf p})g_0(p)h(k_1,q,p+q),
\label{heqk}
\eeq
with the solution
\beq
h(k_1,q,k_2)=\gamma_{e,0}(k_1,q)g_0(k_2)
\left[t({\bf k}_2)-\frac{c(q)}{1+b(q)}\right].
\eeq
Then, inserting Eqs.(\ref{F1}) and (\ref{ff}) into (\ref{sigf2}), and
making use of Eqs.(\ref{gammaderiv})-(\ref{hequation}), we obtain
\beqn
\delta\Sigma^{\sigma\sigma}_1(k)&=&-\frac{2}{\langle Y^{00} \rangle_{-1}}
\lambda_0 \sum_q t({\bf k+q}) g_0(k+q) \Pi_{00}^{\sigma\sigma}(q)
\nonumber\\
&&+\frac{1}{N}\sum_{p,q}
t({\bf p}) g_0(k+q)) g_0(p+q) g_0(p) \gamma_{e,0}(p,q)\frac{c(q)}{1+b(q)}.
\label{deltasig}
\eeqn

The first term in Eq. (\ref{deltasig}) cancels with a similar contribution
in the total self-energy (\ref{selung5}) 
coming from the terms 
$\Gamma_1\left(\var{4}{\alpha}\var{5}{\alpha};\var{2}{00}\right) 
\Gamma_1\left(\var{6}{\alpha}\var{7}{\sigma};\var{1}{\alpha\sigma}\right)$
and
$\Gamma_1\left(\var{4}{\sigma}\var{5}{\alpha};\var{2}{\sigma\alpha}\right)
\Gamma_1\left(\var{6}{\alpha}\var{7}{\alpha};\var{1}{00}\right)$
once the corresponding explicit expressions Eq.(\ref{Gamma1}) for the charge 
vertex are used.

\end{appendix}

\newpage
\begin{figure}
\centerline{\psfig{figure=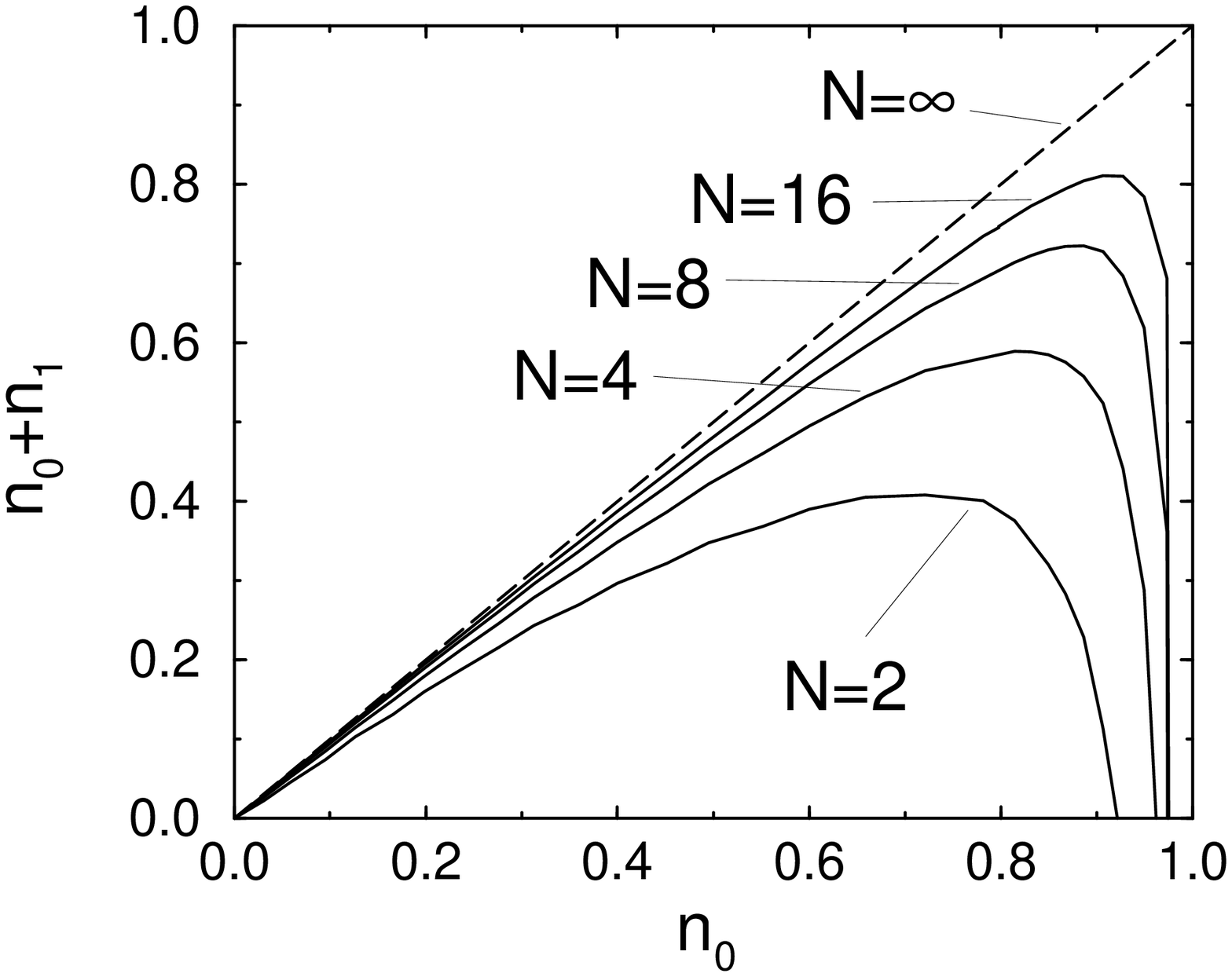,width=15cm}}
\caption{Sum of the $O(1)$ and $O(1/N)$ contributions $n_0$ and
$n_1$, respectively, to the site occupancy  as a function of $n_0$ for
different $N$'s. Only terms which diverge near half-filling haven been
considered in $n_1$.}
\end{figure}
\newpage

\begin{figure}
\centerline{\psfig{figure=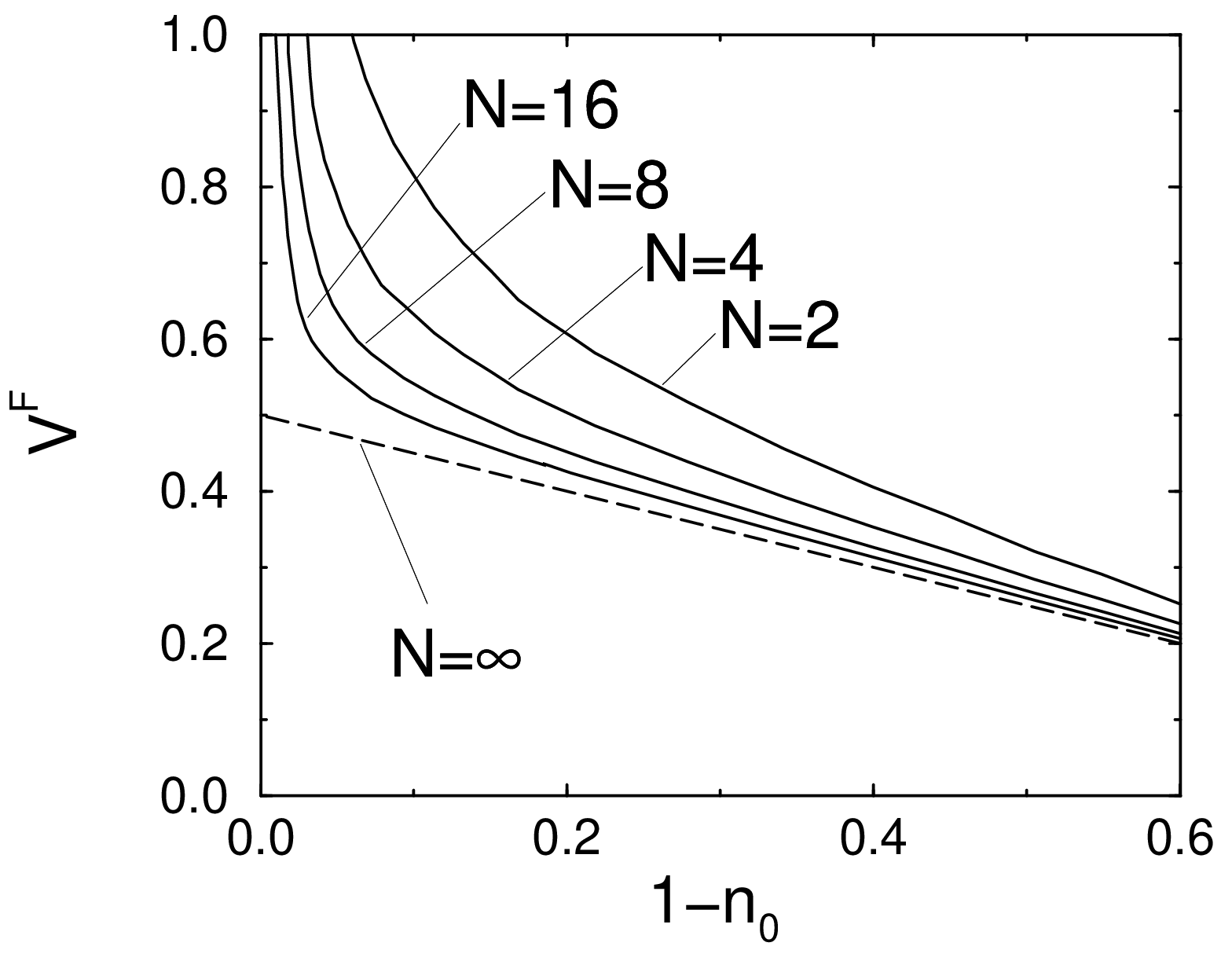,width=15cm}}
\caption{Sum of the $O(1)$ and $O(1/N)$ contributions to the Luttinger
volume $V^F$ as a function of the doping $\delta = 1 - n_0$ keeping
only divergent terms in the $O(1/N)$ contribution. The dashed line 
corresponds to $N=\infty$ and obeys Luttinger's theorem.}
\end{figure}

\end{document}